\documentclass[aps,prl,twocolumn,showpacs,superscriptaddress,groupedaddress,reprint,floatfix]{revtex4}  

\usepackage{graphicx}  
\usepackage{dcolumn}   
\usepackage{bm}        
\usepackage{amssymb}   
\usepackage{color}
\usepackage{hyperref}

\begin{document}

\title{Heterogeneous and anisotropic dynamics of a 2D gel}
\author{D. Orsi}
\affiliation{Physics Department, Parma University, I-43124, Parma, Italy}
\author{L. Cristofolini}
\email{luigi.cristofolini@fis.unipr.it}
\affiliation{Physics Department, Parma University, I-43124, Parma, Italy}
\author{G. Baldi}
\affiliation{Physics Department, Parma University, I-43124, Parma, Italy}
\affiliation{CNR-IMEM Institute, I-43124 Parma, Italy}
\author{A. Madsen}
\affiliation{European Synchrotron Radiation Facility, B.P. 220, F-38043 Grenoble, France}
\affiliation{European X-Ray Free Electron Laser, Albert-Einstein-Ring 19, D-22761 Hamburg, Germany}
\date{\today}

\begin{abstract}We report X-ray Photon Correlation Spectroscopy (XPCS) results on bi-dimensional (2D) gels formed by a Langmuir monolayer of gold nanoparticles. The system allows an experimental determination of the fourth order time correlation function, which is compared to the usual second order correlation function and to the mechanical response measured on macroscopic scale. The observed dynamics is anisotropic, heterogeneous and super-diffusive on the nanoscale. Different time scales, associated with fast heterogeneous dynamics inside 2D cages and slower motion of larger parts of the film, can be identified from the correlation functions. The results are discussed in view of other experimental results and models of three-dimensional gel dynamics.
\end{abstract}
\pacs{64.70.pv, 68.18.-g, 83.80.Kn, 83.85.Hf}
\maketitle

Many of the diverse properties of soft materials originate from their complex structures and dynamics characterized by multiple length and time scales. A wide variety of technologies, from paints to food science and from oil recovery to personal care products, depend crucially on understanding the dynamics of complex interfacial systems. Gels and their dynamics have been widely studied in 3D bulk configurations~\cite{Cipelletti,Anderson,Lu}, however, dimensionality is known to be quantitatively important in arrested systems: e.g. hard spheres caging occurs in 3D at a volume fraction $\Phi_{RCP}^{3D}\simeq60-64\%$, while in 2D the random close packing is attained at an area fraction  $\Phi_{RCP}^{2D}\simeq82\%$~\cite{Berryman}.

The effect of dimensionality on the glass transition process in frictionless colloids has been investigated in numerical simulations~\cite{OHern}, within the mode coupling approach~\cite{Bayer}, and in experiments on granular materials~\cite{Keys,Dauchot}. Heterogeneous dynamics of micron sized beads confined in 2D and interacting through complex inter-particle potentials, either attractive or repulsive, have recently been studied by means of video microscopy and particle tracking techniques~\cite{Yunker,zhang}. Both simulations and theory suggest that the behavior is not significantly different in 2D and in 3D. This is surprising as dimensionality usually plays a decisive role in phase transitions. However, this point still lacks a definite experimental confirmation and, as a matter of fact, a recent comparison between surface and bulk dynamics in a repulsive colloidal suspension evidenced heterogeneous ballistic motion parallel to the surface while the dynamics in the direction perpendicular to the surface was much slower and apparently diffusive \cite{Duri}.

Here we report the first study by X-ray Photon Correlation Spectroscopy (XPCS)~\cite{Grubel} of interfacial dynamics in a strictly two-dimensional system formed by a Langmuir monolayer of weakly attractive nano-particles at the air/water interface. We compare the microscopic dynamics probed with XPCS and the mechanical moduli measured by Interfacial Shear Rheometry (ISR, see refs. \cite{Brooks,Orsi}) on the same samples. Typically, dynamical heterogeneity is identified by calculating the variance $\chi$ of the two-times correlation function $C(t_{1},t_{2})$.  Here, we provide the first experimental determination -- by XPCS -- of the fourth order correlation function $g^{(4)}(t,\widetilde{\tau})$ \cite{Maccarone,Madsen,Glotzer} giving direct information about the lifetime of dynamical heterogeneity upon approaching dynamical arrest.

\begin{figure}
	\setlength\fboxsep{0pt}
	\fbox{\resizebox{0.7\columnwidth}{!}{\includegraphics[clip,trim=0pt 0pt 5pt 5pt]{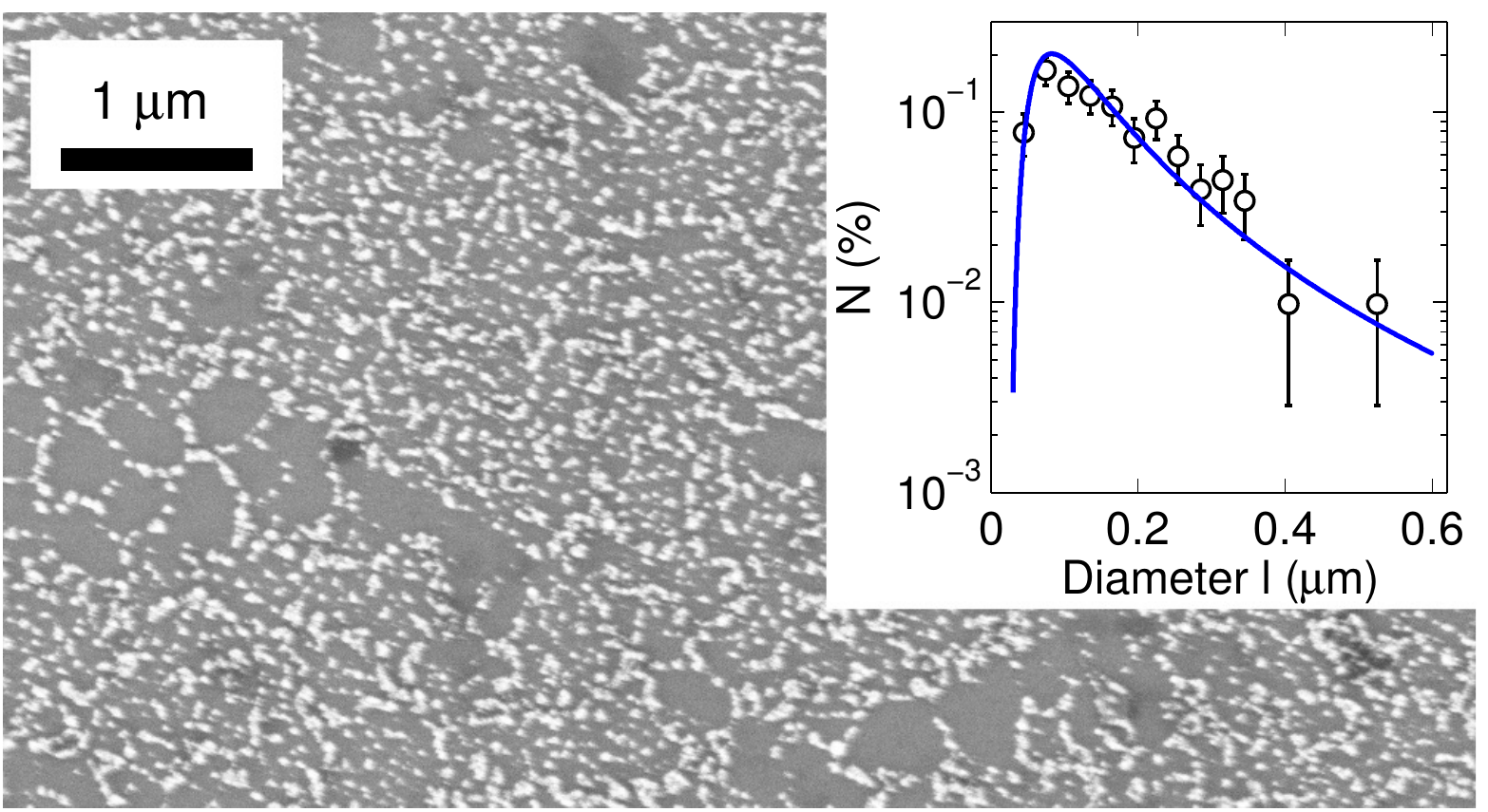}}}\vspace{5pt}
	{\resizebox{0.7\columnwidth}{!}{\includegraphics{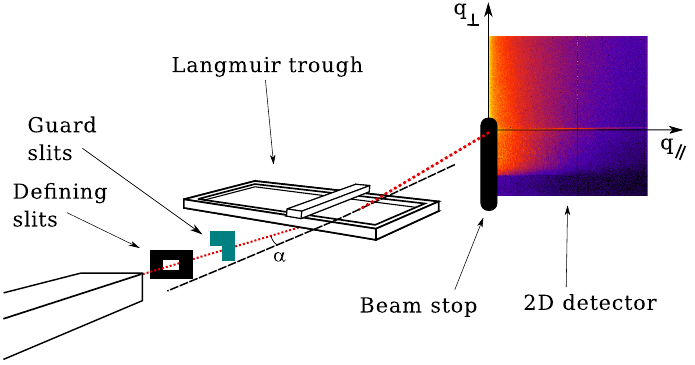}}}
	\caption{ (color online) \textbf{Top}: a typical SEM image of a monolayer with $\Phi=50\%$ transferred onto a solid substrate. \textbf{Inset}: distribution of holes on a log scale, and its best fit with the Levy distribution (continuous line). \textbf{Bottom}: sketch of the setup used for XPCS experiment on a Langmuir monolayer.}
	\label{fig:1}
\end{figure}

Gold  nanoparticles of $70$\AA$ $  diameter, stabilized by dodecanethiol coating, were produced by Ruggeri and coworkers (Pisa University) following the literature \cite{Chen}, and thoroughly characterized by dynamic light scattering and electron microscopy, both in transmission and in scanning (SEM) geometry.  At the air/water interface the particles possess a weakly attractive interaction potential due to hydrophobic interactions. Uniform Langmuir monolayers are prepared through successive slow compression/expansion cycles at constant temperature, following the method given in ref.~\cite{Chen1}. The fraction ($\Phi$) of surface covered by the film is usually deduced from the dispersed aliquot. The determination of $\Phi$ is crucial \cite{PoonWeeksRefA}, therefore in this work it was double checked by in-situ null-ellipsometry and by SEM imaging of monolayers transferred onto a solid substrate. The three independent determinations of $\Phi$ coincide, as reported in \cite{COLSUA}, within an errorbar of roughly $\pm 4 \% $.
A typical SEM image is shown in the top panel of figure \ref{fig:1}. It evidences structural heterogeneity, with clusters and holes of different sizes ranging from a few tens of nm to several microns. The distribution of hole sizes, shown in the inset, is strongly asymmetrical and compatible with a Levy distribution \cite{Gnedenko} characterized by an algebraic tail with exponent $-3$. No crystalline phase was detected neither by X-ray diffraction nor by SEM. The measurements and sample preparation was always performed following the same protocol in order to eliminate uncertainties due to different states of sample aging.

The mechanical properties were characterized on the macro-scale by ISR \cite{Brooks}. As commonly found in many gel systems the loss factor $\tan \delta= G"/G'$ is constant over the measured concentration range, while the complex shear modulus $G^*$ rapidly increases with increasing concentration, and can be modeled as a power law $G^* \propto \left ( \frac{\Phi-\Phi_0 }{\Phi_0} \right )^z$ \cite{stau82, wint86,COLSUA}. In the present case  $\Phi_0=0.27(1)$, close to the concentration at which the gel forms, and the exponent is $z=0.65(1)$, which is smaller than 2.4 observed in protein gels \cite{corr09} and 4.0 found in carbon black gels \cite{tra01}. The difference is probably due to the reduced dimensionality of the present system.

XPCS experiments were performed using partially coherent X-rays at the Tr\"oika beamline ID10A of the European Synchrotron Radiation Facility. A single bounce Si(111) crystal monochromator is used to select 8.06 keV radiation and the transversely coherent beam is defined by slit blades with carefully polished cylindrical edges. A set of guard slits placed just upstream of the sample blocks parasitic scattering due to diffraction from the beam-defining slits. A MAXIPIX detector \cite{Ponchut} is used to collect speckle images close to the specular reflected beam; expressed as a function of $q$, the range from $0.002 $\AA${}^{-1}$ to $0.03 $\AA${}^{-1}$ was covered. Focusing on the slow dynamics, speckle patterns were collected at time intervals of $0.01-0.2 s$. The measurements were performed at constant temperature ($18^\circ$C) and increasing surface pressure, ranging from 10 mN/m up to 30 mN/m, by means of the apparatus sketched in figure \ref{fig:1}. The signals from individual detector pixels were grouped in the analysis in order to provide a $q$-sensitive map of the dynamics and increase the signal-to-noise ratio.

\begin{figure}
	\resizebox{0.225\textwidth}{!}{\includegraphics{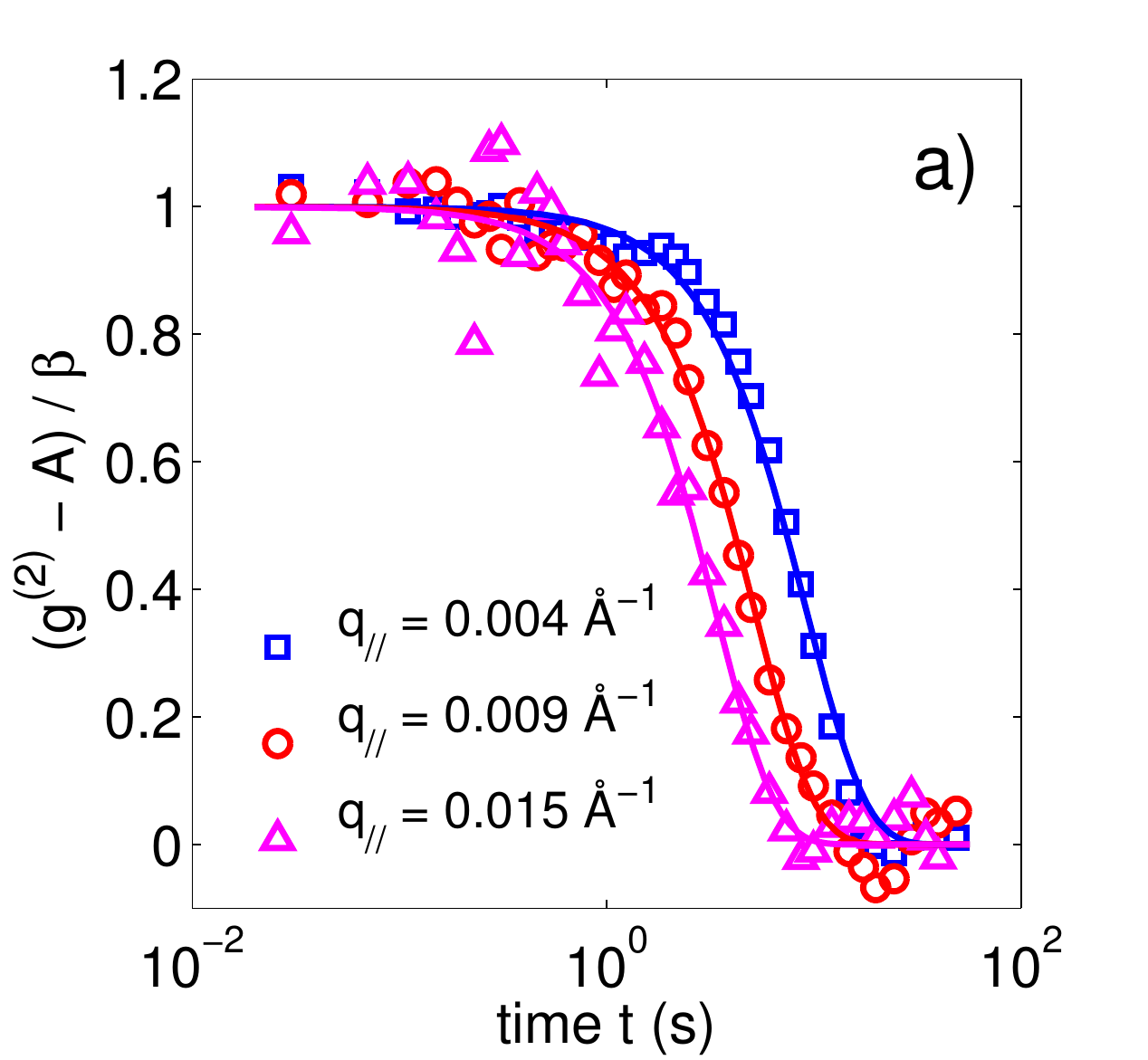}}
	\resizebox{0.225\textwidth}{!}{\includegraphics{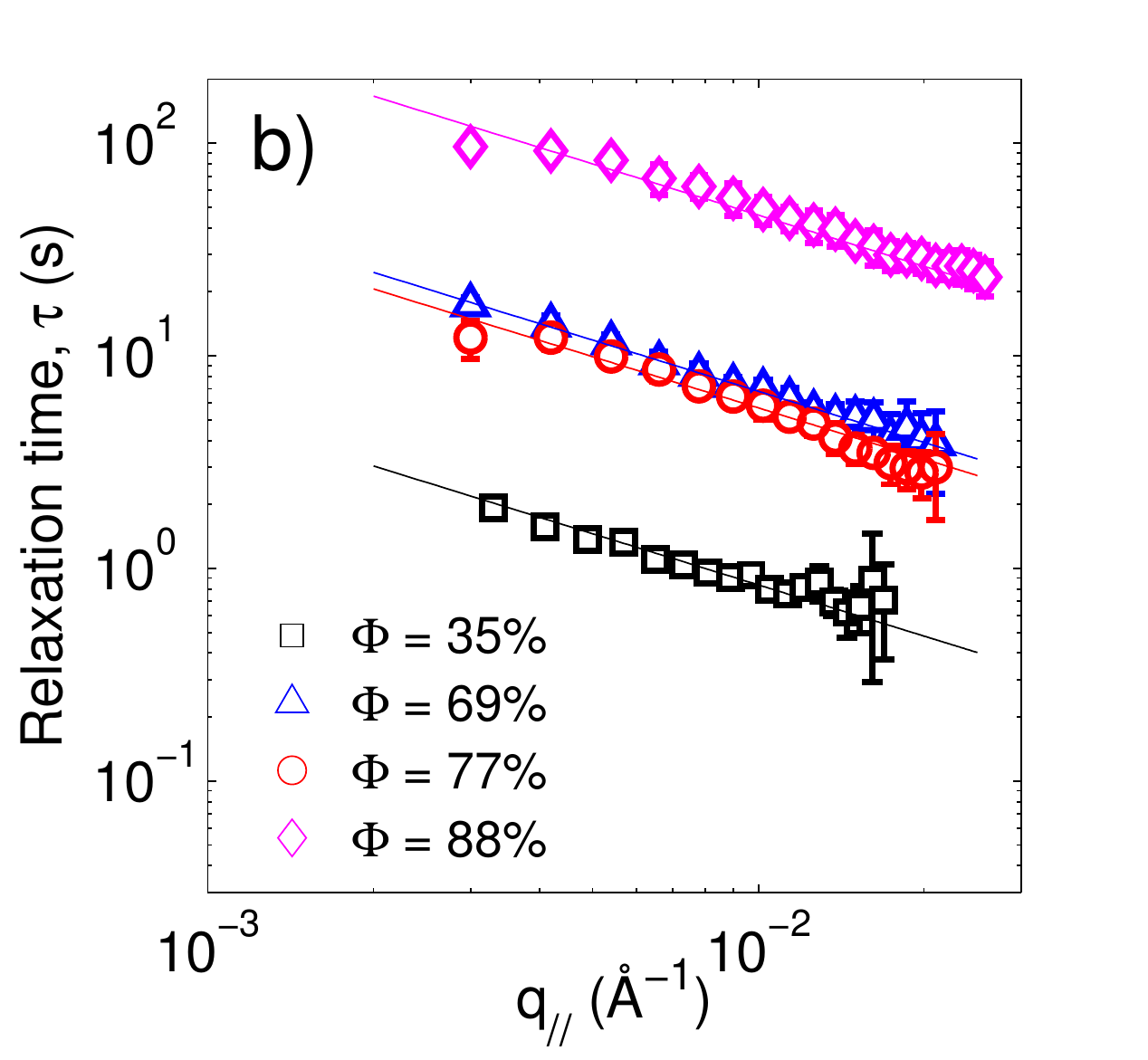}}
	\resizebox{0.5\textwidth}{!}{\includegraphics{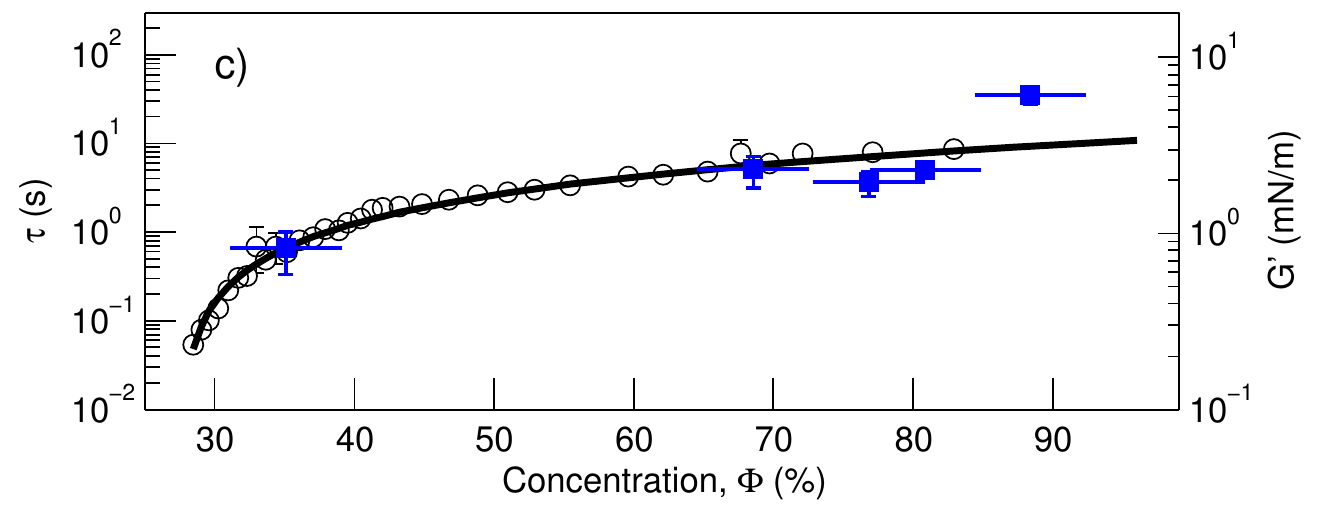}}
\caption{(color online) \textbf{a}: scaled correlation functions $g^{(2)}$ for selected ${q}_{||}$ values (lines are fit to the data) from a Langmuir film at  $\Phi=69\%$.
\textbf{b}: relaxation times $\tau$ for different surface concentrations $\Phi$. Continuous lines are fits to the power law discussed in the text.
\textbf{c}: relaxation times $\tau$ versus surface concentration $\Phi$ at $q=0.015 $\AA${}^{-1}$ (squares, left scale) and shear modulus $G'$ measured by ISR (circles, right scale). The continuous line is the power law fit described in the text. }
	\label{fig:2}
\end{figure}

Intensity auto correlation functions are calculated in the usual way: $g^{(2)}\ \left(q,t\right)={{\frac{\left\langle I(q,t_1)I(q,t_1+t)\right\rangle \ }{{\left\langle I(q,t_1)\right\rangle }^2}}}$ and are reported in figure \ref{fig:2}a for  $\Phi=69\%$ and $q_\perp =0.005$\AA${}^{-1}$. They can be modeled as
\begin{equation}
g^{\left(2\right)}\left(q,t\right)=A+\beta e^{-2{\left({t}/{\tau }\right)}^{\gamma }}
\label{eq:1}
\end{equation}
where $A$ is the baseline, $\beta$ is the contrast, $\tau$ the relaxation time and $\gamma$ the compression coefficient, determining the shape of the relaxation. The shape can be taken as a signature of the interactions present in the system: {\em e.g.} one expects $\gamma=1$ for Brownian motion while $\gamma=1.5$ is often found in colloidal gels \cite{Fluerasu}. We find $\gamma=1.5(1)$ for all the concentrations studied here.
Figure \ref{fig:2}b shows the $q$ dependence of the relaxation times $\tau$ obtained from fits of eq. (\ref{eq:1}) to the correlation functions taken at various concentrations $\Phi$. Different kinds of dynamics can usually be distinguished by the relation between $\tau$ and $q$: in the present case, at all concentrations, a power law $\tau \approx q^{-n}$ is found with exponent $n\simeq0.8(1)$ (continuous lines in the figure). A similar behavior has been found in a variety of arrested systems \cite{Cipelletti}, e.g. in oil nanoemulsion \cite{Guo1}, in aqueous colloidal polystyrene gels \cite{Cipelletti1}, and recently also by us in molecular layers of polymers \cite{Orsi1}.
In figure \ref{fig:2}c the dependence of $\tau$ on $\Phi$ is shown, measured at constant momentum transfer ($q=0.015$ \AA${}^{-1}$), and compared with the 2D storage shear modulus $G'$ measured by ISR on the same film: they share roughly the same power law dependence on $\Phi$ represented by the continuous line.

The relation between microscopic fluctuations ($\tau$) and a macroscopic response function ($G'$) was predicted within the Bouchaud-Pitard model~\cite{Bouchaud}. This model assumes the dynamics to be governed by micro-collapses of the gel, resulting in the formation of force dipoles, which affect the surrounding particles by inducing a strain field. With the further assumptions that the collapses occur randomly in space and time, the stochastic equations of motion for the strain field can be solved analytically. In the slow-collapse regime, the resulting intermediate scattering function has a compressed shape with $\gamma=1.5$, while the relaxation time is proportional to the macroscopic elastic modulus of the gel and inversely proportional to the momentum $q$. Our findings confirm all the model predictions.

\begin{figure}
	\resizebox{0.48\columnwidth}{!}{\includegraphics{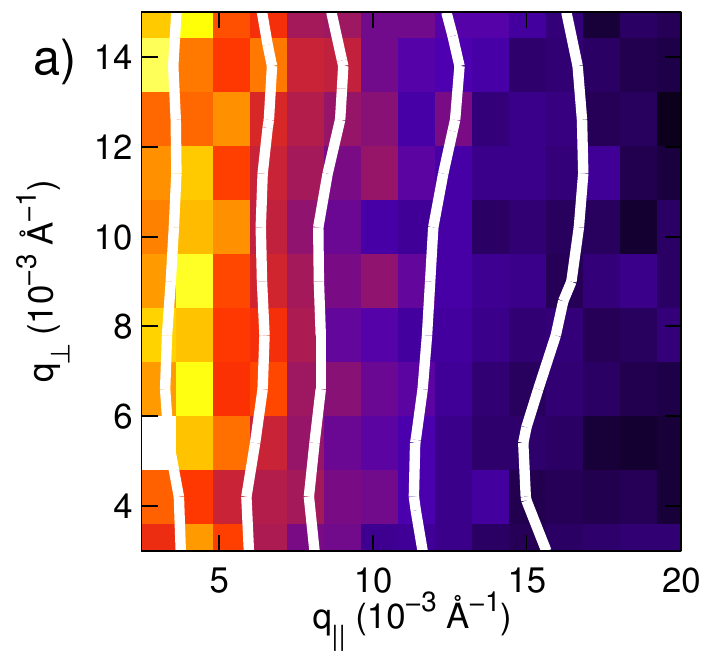}}
	\resizebox{0.48\columnwidth}{!}{\includegraphics{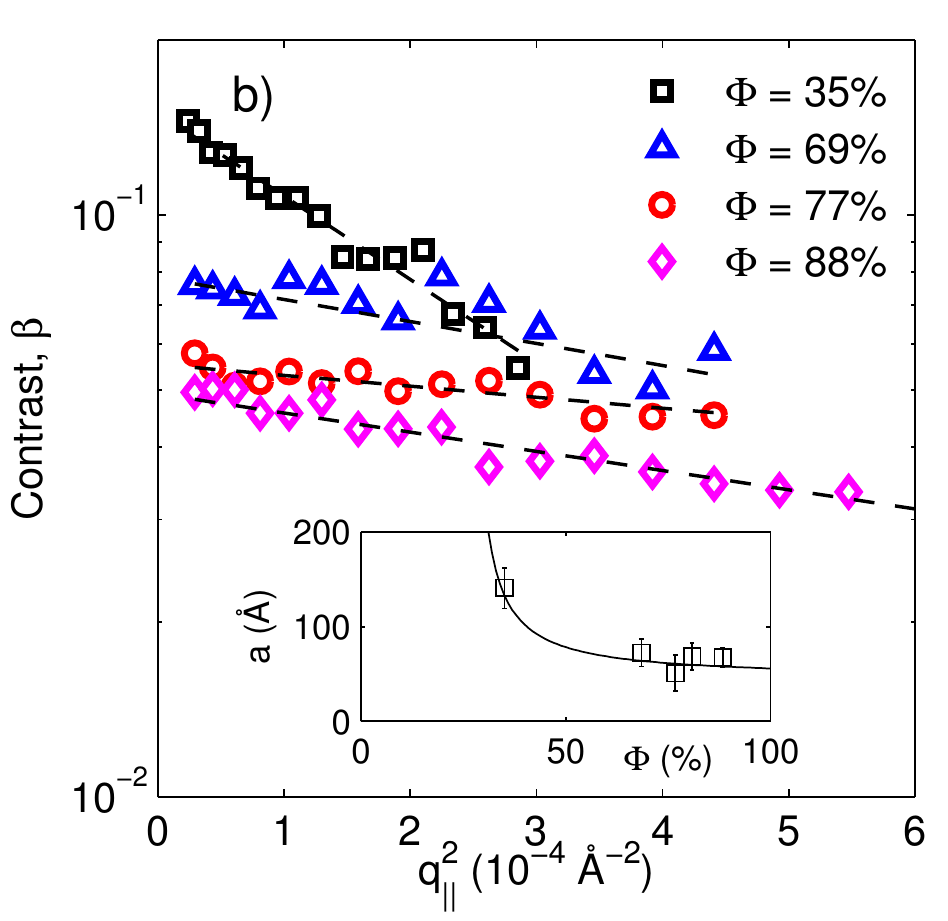}}
	\resizebox{0.49\columnwidth}{!}{\includegraphics{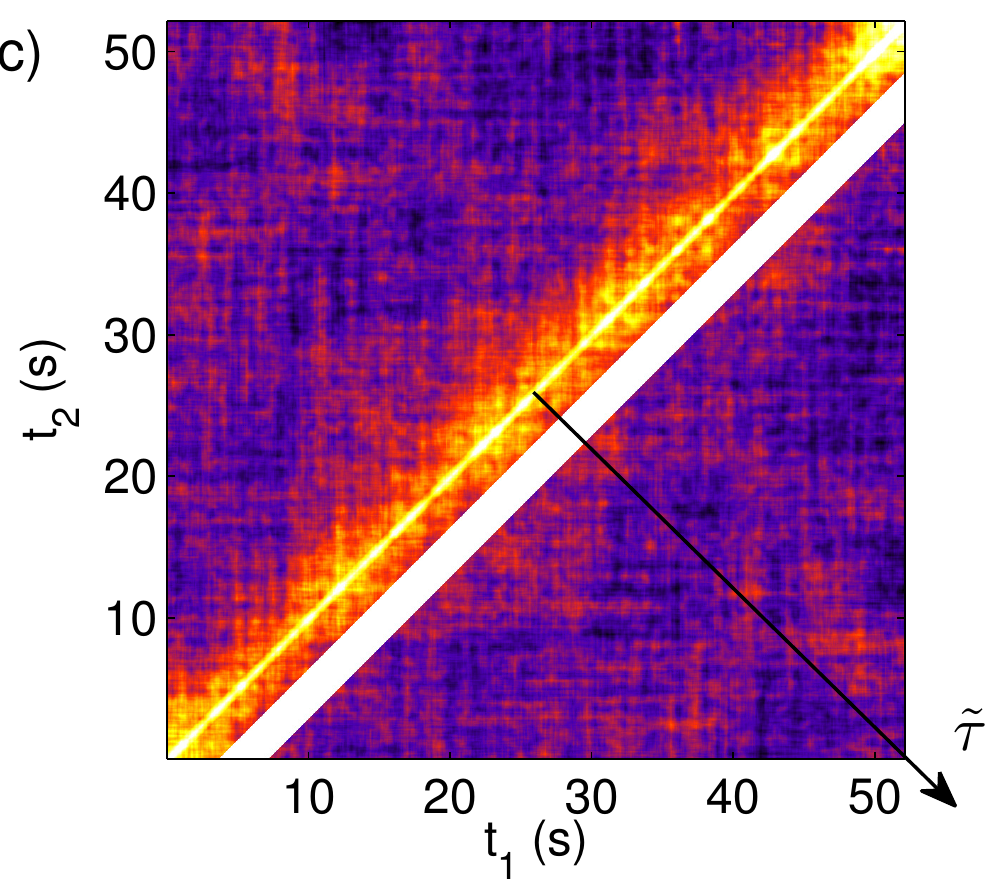}}
	\resizebox{0.47\columnwidth}{!}{\includegraphics{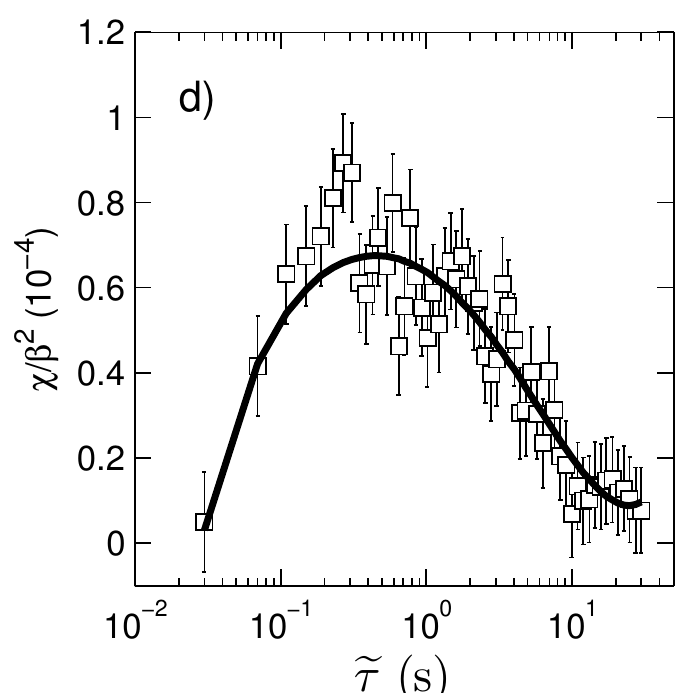}}
	\caption{(color online) \textbf{a}: $q$-map of relaxation times $\tau$ measured at  $\Phi=69\%$ along different directions in reciprocal space: ${q}_{||}$ and ${q}_{\perp}$ are respectively parallel and perpendicular to the water surface. Solid lines are isochrones.
\textbf{b}: contrast $\beta$ as a function of squared momentum ${q}_{||}^{2}$, for different values of $\Phi$. Dashed lines are Debye-Waller fits. \textbf{Inset}: localization length at different values of $\Phi$. The continuous line is a guide to the eye.
\textbf{c:} two times correlation function for  $\Phi=69\%$, $T=18^\circ C$ and ${q}_{||}=0.009$\AA${}^{-1}$. The lighter stripe represents the range over which  $g^{(4)}$ is calculated, as described in the text. The arrow indicates the time $\widetilde \tau$ of equation (\ref{defchi4}).
\textbf{d:} normalized variance $\chi$ of the two-times correlation function reported in panel c; the continuous line is a guide to the eye.}
	\label{fig:3}
\end{figure}

Figure \ref{fig:3}a displays the $q$-map of the relaxation times $\tau$ obtained by fitting $g^{(2)}$ with eq. (\ref{eq:1}). The dynamics is clearly hallmarked by the anisotropic dependence of $\tau$ on the exchanged momentum, where $\tau$ depends only on the surface parallel component ${q}_{||}$ (horizontal direction in the map) and not on the vertical component ${q}_{\perp}$. This picture holds for all the concentrations studied, proving that the nanoparticle dynamics is confined within the air/water interfacial plane.

The correlation functions characterize dynamics within a given time window, whose short-time limit is mainly determined by the amount of scattered intensity and by the detector speed. Any faster dynamics will reduce the contrast of the correlation function to less than the Siegert factor ($\sim$ 0.2 in the present setup). Information on fast motion, if present, can be obtained by studying the missing contrast and its q-dependence. In this case the contrast $\beta$ follows a pseudo Debye-Waller law \cite{Guo11,Bandyo,Czakkel}:
$\beta ={\beta }_0{\rm \ exp}\left(-\frac{q^2a^2}{3}\right),$
indicating that at least parts of the missing contrast is due to a faster rattling motion characterized by the localization length $a$. The fits are displayed in figure \ref{fig:3}b. The inset shows the variation of $a$: with growing $\Phi$ it decreases to reach a limiting value of $50$ \AA, comparable to the mean particle radius. The extrapolation of $\beta$ to $q_{||}=0$ never reaches the Siegert factor, suggesting that several fast dynamical modes contribute to reducing the contrast. For instance, more mobile particles with mean square displacements larger than $1/q_{min}\simeq 250$\AA \cite{Guo11} could be present.

With several populations of particles moving distinctly, the presence of dynamical heterogeneity in the fast dynamics is implied and can be further characterized by higher order correlation functions. In figure \ref{fig:3}c we report the normalized two times correlation function $C(t_{1},t_{2})= I(t_{1})*I(t_{2})$ \cite{Malik} measured at $q_{||}=0.009$\AA${}^{-1}$, $T= 18^\circ C$ and  $\Phi=69\%$. Its variance $\chi$, calculated as in ref.~\cite{Duri1}, is shown in figure \ref{fig:3}d: it features a broad maximum around $\tau_{c}=0.5-1s$, which is an evidence of dynamical heterogeneity. Thanks to the good quality of the data it is possible to bring the analysis one step further and calculate the four times correlation function  $g^{(4)}$ defined as:
\begin{eqnarray}
g^{(4)}\left(t,\widetilde{\tau }\right)={\left\langle C\left(t_1,\ t_1+\widetilde{\tau }\right)\cdot C\left(t_1+t,\ t_1+t+\widetilde{\tau }\right)\right\rangle }_{t_1}= \nonumber \\
{\left\langle I\left(t_1\right)\cdot I\left(\ t_1+\widetilde{\tau }\right)\cdot I\left(t_1+t\right)\cdot I\left(\ t_1+t+\widetilde{\tau }\right)\right\rangle }_{t_1}\
\label{defchi4}
\end{eqnarray}

The first lag time $\widetilde{\tau }=|t_1-t_2|$,  indicated by the lighter stripe in figure \ref{fig:3}c, identifies the sub-diagonal of matrix $C$ along which  $g^{(4)}$ is calculated. The second lag time, $t$, corresponds to the separation between two instants of the selected sub-diagonal. Averages are performed over initial time $t_{1}$. We choose $\widetilde{\tau }={\tau }_c$ because the variance analysis suggests that on this time scale the heterogeneous dynamics is most pronounced.

\begin{figure}
	\resizebox{0.525\columnwidth}{!}{\includegraphics{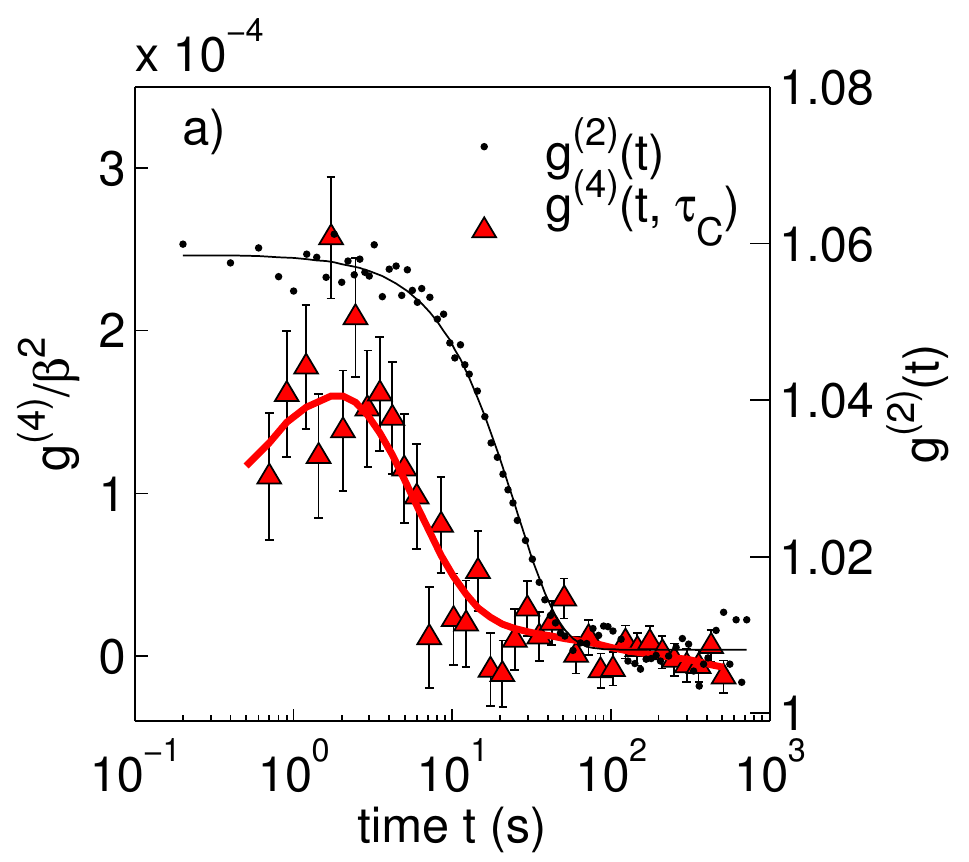}}
	\resizebox{0.46\columnwidth}{!}{\includegraphics{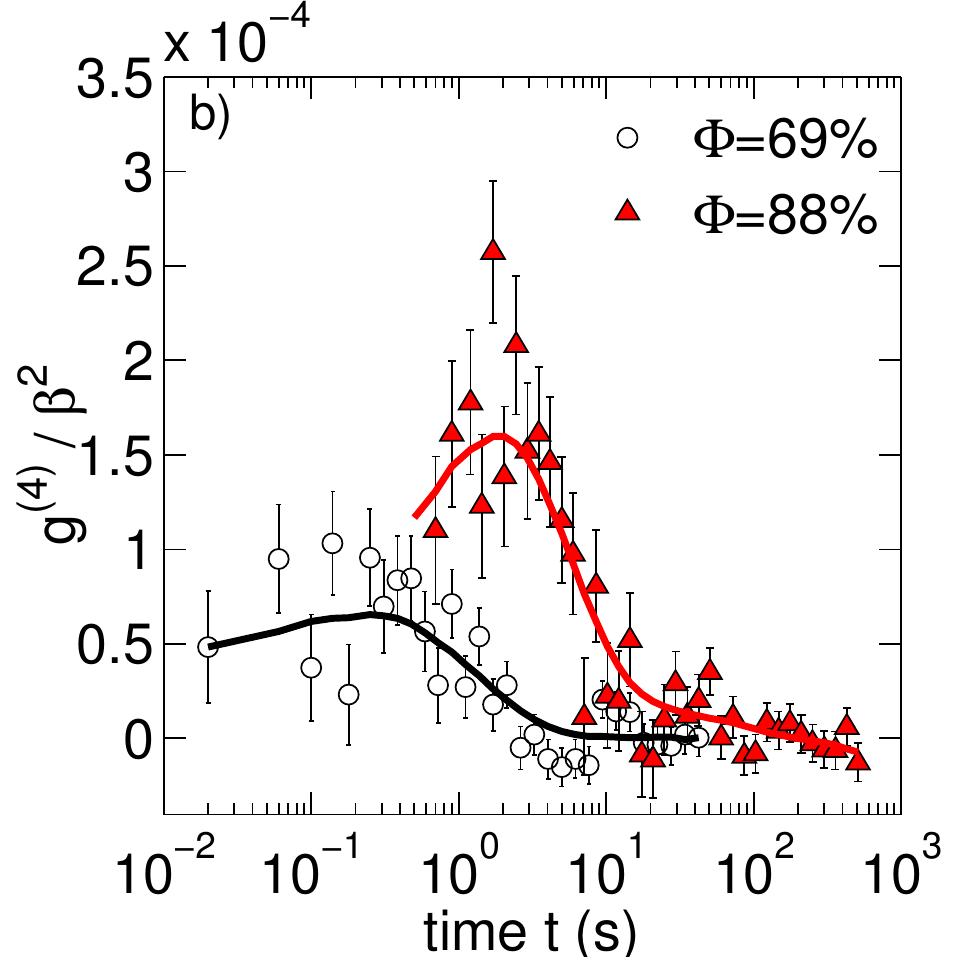}}
	\resizebox{0.48\columnwidth}{!}{\includegraphics{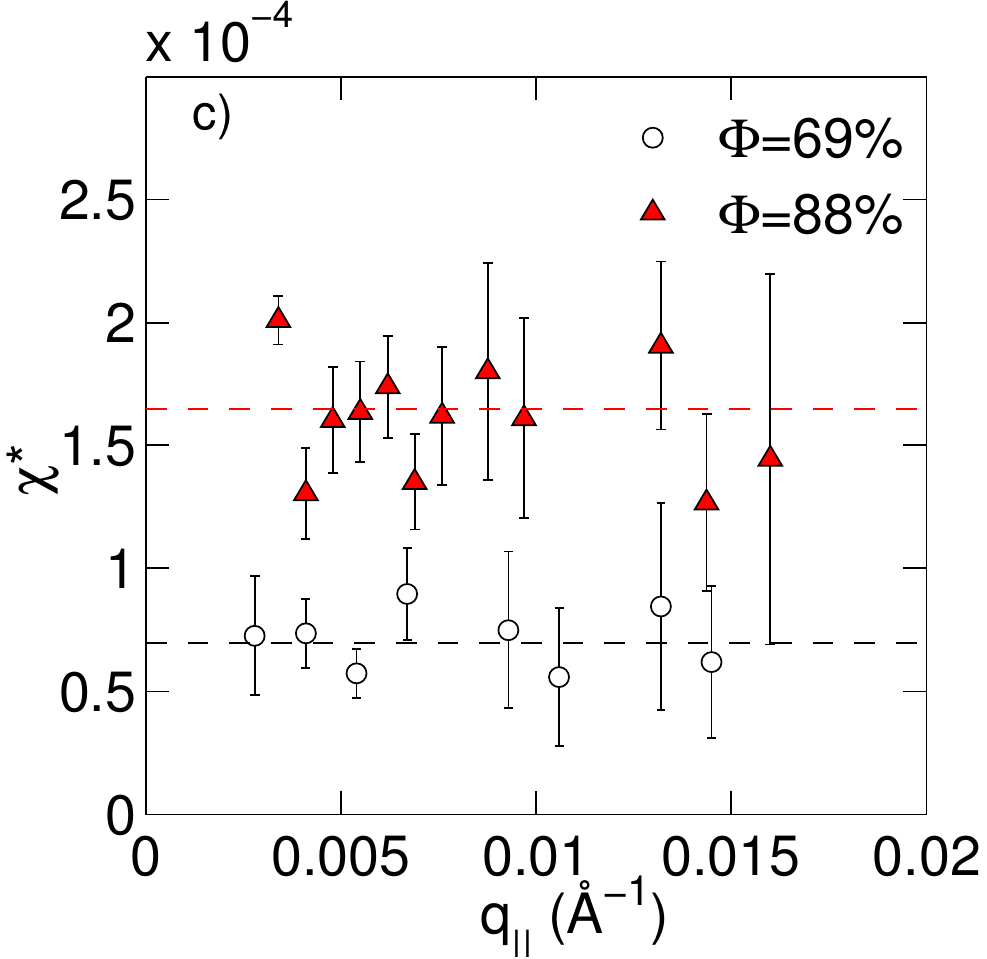}}\hspace{7pt}
	\resizebox{0.46\columnwidth}{!}{\includegraphics{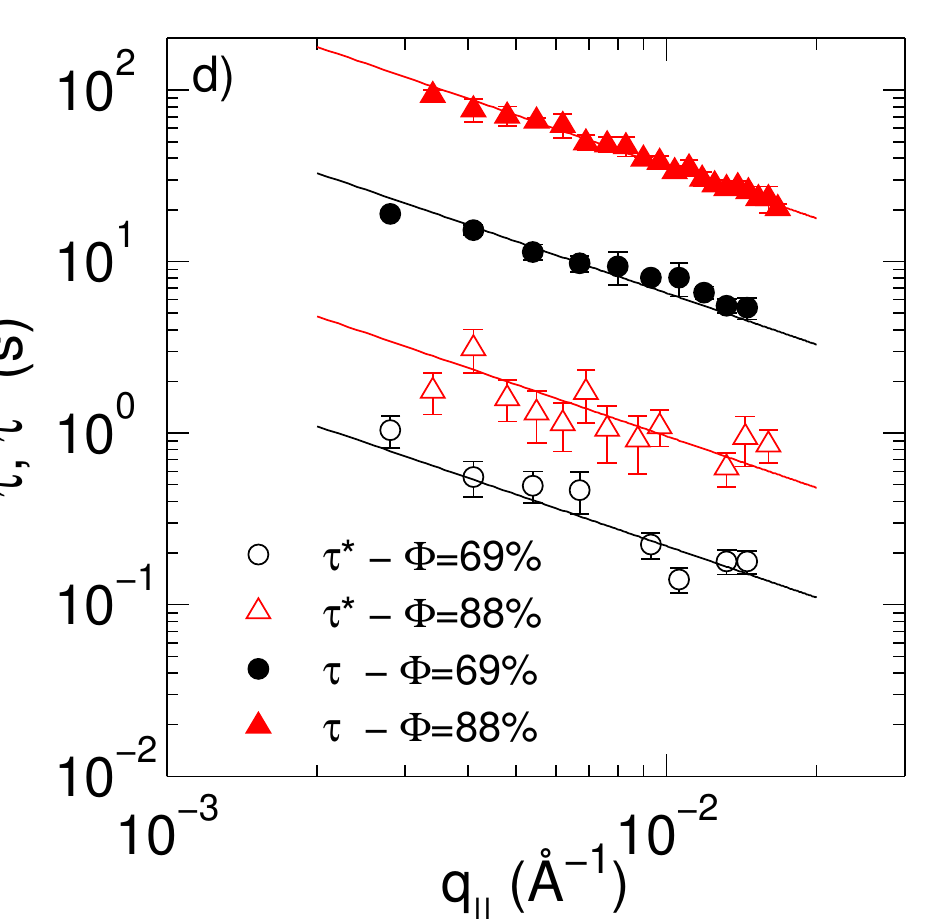}}
\caption{(color online)
\textbf{a:} comparison between $g^{(4)}$ and $g^{(2)}$ both measured at $\Phi=88\%$ and ${q}_{||}=0.009$\AA$^{-1}$. Lines are guides to the eye.
 \textbf{b:} normalized  $g^{(4)}$ for different values of $\Phi$ and ${q}_{||}=0.009$\AA$^{-1}$. Lines are guides to the eye.
\textbf{c:} Intensity $\chi^*$ of the peak of $g^{(4)}$ as a function of $q_{||}$, calculated for $\Phi=69\%$ and $\Phi=88\%$; dashed lines represent their averages.
\textbf{d:} $q_{||}$-dependence of characteristic times $\tau^*$(empty symbols) and of $\tau$ (filled symbols) for the values of $\Phi $ indicated.  They all follow an approximate $\sim {q}_{||}^{-1}$ scaling law (lines).}
\label{fig:4}
\end{figure}

Figure \ref{fig:4}a reports $g^{(4)}$ , for ${q}_{||}=0.009$\AA${}^{-1}$ and $\widetilde{\tau }={\tau }_c$, and compares it with the usual 2nd order correlation function $g^{(2)}$. Both functions display a full decay during the timescales accessible in the experiment but the functional form of $g^{(4)}$ appears to be characterized by a peak located at $\tau^{*}$. This observation holds for all concentrations studied, as shown in \ref{fig:4}b. We take $\tau^*$ as a measure of the timescale of dynamical heterogeneity and track its dependence on $q$ as shown in \ref{fig:4}d. In accordance with the previous discussion concerning the "missing contrast" it is found that $\tau^{*} < \tau$ for all $q$ and for all concentrations studied. In addition the $q$-dependencies of $\tau^*$ and $\tau$ appear to be similar ($\sim {q}_{||}^{-1}$)  even if the two time scales originate from different dynamics. Although akin observations have been made before, see e.g. Ref.~\cite{Maccarone}, a deeper understanding of this "scaling behavior" is still missing. No ${q}_{\perp}$ dependence of the variance $\chi$ or of ${\tau }^*$ was observed, indicating that all the observed dynamics (fast and slow) and the heterogeneous behavior are surface phenomena of 2D nature.

For the highest concentration $g^{(4)}$ has a larger amplitude ($\chi^*$) as observed in \ref{fig:4}b, which in turn appears at slower times. In the analogy between temperature and packing fraction, this behavior is in accordance with predictions of numerical simulations \cite{Berthier}. A similar behavior was also found in the variance $\chi$ measured by XPCS \cite{Madsen} and photon correlation imaging \cite{Dur09} in 3D systems. Here, the decreased mobility is followed by an increase in the range of spatial and temporal correlations between particles, in agreement with the general assumption that dynamical arrest is related to the development of heterogeneities.

Within the statistical uncertainty, no clear $q$-dependence could be detected in $\chi^*$, as shown in figure \ref{fig:4}c. This suggests that a hierarchy of different length scales is involved in the heterogeneous behavior, and possibly the fast, non-localized population could be modeled as Levy flights \cite{Buonsante,lilliput}, similarly to other systems presenting dynamical heterogeneity \cite{Maccarone,Orsi1}.

In conclusion, we present the first XPCS experiment on a Langmuir monolayer (2D gel) displaying anisotropic and heterogeneous dynamical behavior. The data quality allows a rare experimental characterization of the dynamics by calculating the fourth order temporal correlation function $g^{(4)}$. The anisotropic dynamics occurs on different time -- and length -- scales: fast collective heterogeneous rearrangements characterized by $\tau^*$ are responsible for the peak in $g^{(4)}$, but are too fast to be directly seen in ${g}^{(2)}$, where they only appear as reductions of the contrast. Part of the missing contrast can be accounted for using a Debye-Waller model to describe fast rattling on the nano scale but the data also indicate that fast diffusive-like motion is present. $g^{(2)}$ tracks a much slower ergodicity restoring dynamics characterized by the relaxation time $\tau$. Both fast and slow motions are anisotropic and confined to the surface plane (2D) and have similar dependencies on $q$, and on the coverage fraction $\Phi$. Finally, we note that the relaxation time scales with $\Phi$ similarly to the 2D mechanical response measured by ISR. This, together with the non-Brownian character of the ergodicity restoring dynamics, is in good agreement with the model proposed by Bouchaud and Pitard \cite{Bouchaud}.

\begin{acknowledgements}
The ESRF is acknowledged for provision of beam time, O. Konovalov (ESRF) for access to the Langmuir through, G. Ruggeri and A. Pucci  (Pisa) for preparing the nanoparticles, F. Pincella (Parma) for help with the experiments, R. Leheny (John Hopkins) and P. Cicuta (Cambridge) for useful comments and discussions.

\end{acknowledgements}

\end{document}